\documentclass[nofootinbib,prd,aps,superscriptaddress,preprintnumbers,twocolumn,showpacs]{revtex4-1}
\usepackage[dvipdfmx]{graphicx}
\usepackage{amsmath,amssymb,amsthm,bm,bigdelim,multirow}
\usepackage{color}
\usepackage{booktabs}
\usepackage{braket,cases,latexsym}
\usepackage{here}

\begin{document}
\title{Tensor renormalization group approach to (1+1)-dimensional Hubbard model}

  \author{Shinichiro Akiyama}
  \email[]{akiyama@het.ph.tsukuba.ac.jp}
  \affiliation{Graduate School of Pure and Applied Sciences, University of Tsukuba, Tsukuba, Ibaraki
    305-8571, Japan}

  \author{Yoshinobu Kuramashi}
  \email[]{kuramasi@het.ph.tsukuba.ac.jp}
  \affiliation{Center for Computational Sciences, University of Tsukuba, Tsukuba, Ibaraki
    305-8577, Japan}
    
\begin{abstract}
We investigate the metal-insulator transition of the (1+1)-dimensional Hubbard model in the path-integral formalism with the tensor renormalization group method. The critical chemical potential $\mu_{\rm c}$ and the critical exponent $\nu$ are determined from the $\mu$ dependence of the electron density in the thermodynamic limit. Our results for $\mu_{\rm c}$ and $\nu$ show consistency with an exact solution based on the Bethe ansatz. Our encouraging results indicate the applicability of the tensor renormalization group method to the analysis of higher-dimensional Hubbard models.
\end{abstract}
\date{\today}

\preprint{UTHEP-756, UTCCS-P-137}

\maketitle

\section{Introduction}
\label{sec:intro}

The tensor renormalization group (TRG) method \footnote{In this paper the TRG method or the TRG approach refers to not only the original numerical algorithm proposed by Levin and Nave \cite{Levin:2006jai} but also its extensions \cite{PhysRevB.86.045139,Shimizu:2014uva,Sakai:2017jwp,Adachi:2019paf,Kadoh:2019kqk,Akiyama:2020soe}.}, which was originally proposed to study two-dimensional (2$d$) classical spin systems in the field of condensed matter physics~\cite{Levin:2006jai}, has been now used to study wide varieties of models in particle physics taking several advantages over the Monte Carlo method. (i) The TRG method does not suffer from the sign problem as already confirmed by studying various quantum field theories \cite{Shimizu:2014uva,Shimizu:2014fsa,Shimizu:2017onf,Takeda:2014vwa,Kadoh:2018hqq,Kadoh:2019ube,Kuramashi:2019cgs,Akiyama:2020ntf,Akiyama:2020soe}. (ii) Its computational cost depends on the system size only logarithmically. (iii) It allows direct manipulation of the Grassmann variables \cite{Shimizu:2014uva,Sakai:2017jwp,Yoshimura:2017jpk,Akiyama:2020soe}. (iv) We can obtain the partition function or the path-integral itself.

The sign problem is common both in particle physics and condensed matter physics. A typical example in particle physics is the lattice QCD at finite density, where an introduction of the chemical potential causes the sign problem, and the Hubbard model is notorious in the condensed matter physics.
Recently the authors have successfully applied the TRG method to analyze the phase transition of the 4$d$ Nambu$-$Jona-Lasinio (NJL) model at high density and very low temperature~\cite{Akiyama:2020soe}. The study of the NJL model has two important aspects. Firstly, the NJL model is a prototype of QCD. Their phase structures are expected to be similar so that the study of the NJL model at finite density is a good testbed before investigating the finite density QCD. Secondly, the NJL model has a similar path-integral form to the Hubbard model: Both consist of a hopping term and a four-fermi interaction term. This indicates that the technical details of the TRG method employed in the analysis of the NJL model could apply to the Hubbard model. It is interesting to investigate whether or not the TRG method overcomes the sign problem in the Hubbard model.

In this paper, we investigate the metal-insulator transition of the (1+1)$d$ Hubbard model in the path-integral formalism by calculating the electron density as a function of the chemical potential $\mu$. After examining the imaginary-time discretization effects and the temperature dependence, we determine the critical value of the chemical potential $\mu_{\rm c}$ and the critical exponent $\nu$ in the thermodynamic limit at the zero temperature. Our results for $\mu_c$ and $\nu$ show agreement with the theoretical prediction based on the Bethe ansatz~\cite{PhysRevLett.20.1445,LIEB20031}.

This paper is organized as follows. In Sec.~\ref{sec:method} we define the Hubbard model in the path-integral formalism and explain the numerical algorithm. In Sec.~\ref{sec:results} we show our results and compare them with theoretical predictions.  Section~\ref{sec:summary} is devoted to summary and outlook.

\section{Formulation and numerical algorithm}
\label{sec:method}

\subsection{(1+1)-dimensional Hubbard model in the path-integral formalism}
\label{subsec:action}

We consider the partition function of the Hubbard model in the path-integral formalism on an anisotropic rectangular lattice with the physical volume $V=L\times \beta$, whose spatial extension is defined as $L=aN_{\sigma}$ with $a$ the spatial lattice spacing. $\beta$ denotes the inverse temperature, which is divided as $\beta=1/T=\epsilon N_\tau$.
The path-integral expression of the partition function is given by \footnote{See Ref.~\cite{Creutz:1986ky} or Refs.~\cite{10.2307/2033649,Suzuki:1976be} for the conversion procedure from the operator formalism to the path-integral one.} 
\begin{align}
	Z=\int\left(\prod_{n\in\Lambda_{1+1}}\prod_{s=\uparrow,\downarrow}{\rm d}\bar{\psi}_{s}(n){\rm d}\psi_{s}(n)\right){\rm e}^{-S},
	\label{eq:Z}
\end{align}
where $n=(n_{\sigma},n_{\tau})\in\Lambda_{1+1}(\subset\mathbb{Z}^2)$ specifies a position in the lattice $|\Lambda_{1+1}|=N_\sigma\times N_\tau$. Since the Hubbard model describes the spin-1/2 fermions, they are labeled by $s=\uparrow,\downarrow$, corresponding to the spin-up and spin-down, respectively. Introducing the notation,
\begin{align}
	\psi(n)=\left(
	\begin{array}{c}
	 	\psi_\uparrow(n)\\ \psi_\downarrow(n) 
	\end{array}
	\right),
	~\bar{\psi}(n)=\left(\bar{\psi}_\uparrow(n),\bar{\psi}_\downarrow(n)\right),
\end{align}
the action $S$ is defined as
\begin{widetext}
\begin{align}
\label{eq:action}
        	S&=\sum_{n_\tau,n_{\sigma}}\epsilon a\left\{\bar{\psi}(n)\left(\frac{\psi(n+{\hat \tau})-\psi(n)}{\epsilon}\right)-t\left(\bar{\psi}(n+{\hat\sigma})\psi(n)+\bar{\psi}(n)\psi(n+{\hat\sigma})\right)+\frac{U}{2}\left(\bar{\psi}(n)\psi(n)\right)^2-\mu\bar{\psi}(n)\psi(n)\right\}.
\end{align}
\end{widetext}
The kinetic term in the spatial direction contains the hopping parameter $t$. The four-fermi interaction term represents the Coulomb repulsion of electrons at the same lattice site. The chemical potential is denoted by the parameter $\mu$. Note that the half-filling is realized at $\mu=U/2$ in the current definition. We assume the periodic boundary condition in the spatial direction, $\psi(N_{\sigma}+1,n_\tau)=\psi(1,n_\tau)$, while the anti-periodic one in the temporal direction, $\psi(n_{\sigma},N_\tau+1)=-\psi(n_{\sigma},1)$. In the following discussion, we always set $a=1$.

\subsection{Tensor network representation}
\label{subsec:tn-rep}

Now, we introduce the tensor network representation for Eq.~\eqref{eq:Z}, based on Ref.~\cite{Akiyama:2020sfo}. At each lattice site, we define the Grassmann tensor $\mathcal{T}$ by
\begin{widetext}
\begin{align}
\label{eq:gtensor}
	&\mathcal{T}_{\Psi_{\sigma}\Psi_{\tau}\bar{\Psi}_{\tau}\bar{\Psi}_{\sigma}}=\sum_{i_{\sigma,\uparrow},i_{\sigma,\downarrow},j_{\sigma,\uparrow},j_{\sigma,\downarrow}}~\sum_{i_{\tau,\uparrow},i_{\tau,\downarrow}}~\sum_{i'_{\sigma,\uparrow},i'_{\sigma,\downarrow},j'_{\sigma,\uparrow},j'_{\sigma,\downarrow}}~\sum_{i'_{\tau,\uparrow},i'_{\tau,\downarrow}}\nonumber\\
	&\times T_{(i_{\sigma,\uparrow},i_{\sigma,\downarrow},j_{\sigma,\uparrow},j_{\sigma,\downarrow})(i_{\tau,\uparrow},i_{\tau,\downarrow})(i'_{\sigma,\uparrow},i'_{\sigma,\downarrow},j'_{\sigma,\uparrow},j'_{\sigma,\downarrow})(i'_{\tau,\uparrow},i'_{\tau,\downarrow})}\Psi_{\sigma,1}^{i_{\sigma,\uparrow}}\Psi_{\sigma,2}^{i_{\sigma,\downarrow}}\Psi_{\sigma,3}^{j_{\sigma,\uparrow}}\Psi_{\sigma,4}^{j_{\sigma,\downarrow}}\Psi_{\tau,1}^{i_{\tau,\uparrow}}\Psi_{\tau,2}^{i_{\tau,\downarrow}}\bar{\Psi}_{\tau,2}^{i'_{\tau,\downarrow}}\bar{\Psi}_{\tau,1}^{i'_{\tau,\uparrow}}\bar{\Psi}_{\sigma,4}^{j'_{\sigma,\downarrow}}\bar{\Psi}_{\sigma,3}^{j'_{\sigma,\uparrow}}\bar{\Psi}_{\sigma,2}^{i'_{\sigma,\downarrow}}\bar{\Psi}_{\sigma,1}^{i'_{\sigma,\uparrow}},
\end{align}
\end{widetext}
where $T$ is called the coefficient tensor, whose components are in $\mathbb{R}$ and all the subscripts of the coefficient tensor take 0 or 1. We have introduced the auxiliary Grassmann fields $\Psi_{\sigma}=(\Psi_{\sigma,1},\Psi_{\sigma,2},\Psi_{\sigma,3},\Psi_{\sigma,4}),$ $\bar{\Psi}_{\sigma}=(\bar{\Psi}_{\sigma,4},\bar{\Psi}_{\sigma,3},\bar{\Psi}_{\sigma,2},\bar{\Psi}_{\sigma,1})$, $\Psi_{\tau}=(\Psi_{\tau,1},\Psi_{\tau,2})$, and $\bar{\Psi}_{\tau}=\bar{\Psi}_{\tau,2},\bar{\Psi}_{\tau,1})$. In Eq.~\eqref{eq:action}, we have two types of hopping terms in the spatial direction. On the other hand, we have just one type of hopping in the temporal direction. Since the model describes spin-1/2 particles, the spatial auxiliary Grassmann field $\Psi_{\sigma}$ has 2 (hopping terms) $\times$ 2 (spin d.o.f.) components and  the temporal one $\Psi_{\tau}$ has $1\times2$ components. Using the Grassmann tensor $\mathcal{T}$ in Eq~\eqref{eq:gtensor}, the path integral $Z$ is expressed by
\begin{widetext}
	\begin{align}
\label{eq:gtn}
	Z&=\int\left(\prod_{n\in\Lambda_{1+1}}{\rm d}\bar{\Psi}_{\tau}(n){\rm d}\Psi_{\tau}(n){\rm d}\bar{\Psi}_{\sigma}(n){\rm d}\Psi_{\sigma}(n)~{\rm e}^{-\left(\bar{\Psi}_{\sigma}(n)\Psi_{\sigma}(n)+\bar{\Psi}_{\tau}(n)\Psi_{\tau}(n)\right)}\right)\prod_{n\in\Lambda_{1+1}}\mathcal{T}_{\Psi_{\sigma}(n)\Psi_{\tau}(n)\bar{\Psi}_{\tau}(n-\hat{\tau})\bar{\Psi}_{\sigma}(n-\hat{\sigma})}.
\end{align}
\end{widetext}
See Appendix~\ref{app:gtensor} for the detailed explanation to derive the above Grassmann tensor and its tensor network.

\subsection{Numerical algorithm}
\label{subsec:algorithm}

We employ the higher-order TRG (HOTRG) algorithm \cite{PhysRevB.86.045139} to evaluate the Grassmann tensor network in Eq~\eqref{eq:gtn}. Using the HOTRG, we firstly carry out $m_{\tau}$ times of renormalization along the temporal direction. This procedure converts the initial Grassmann tensor $\mathcal{T}_{\Psi_{\sigma}\Psi_{\tau}\bar{\Psi}_{\tau}\bar{\Psi}_{\sigma}}$ into the coarse-grained one $\mathcal{T}_{\Xi_{\sigma}\Psi_{\tau}\bar{\Psi}_{\tau}\bar{\Xi}_{\sigma}}$. Secondly, we employ the $2d$ HOTRG procedure, regarding $\mathcal{T}_{\Xi_{\sigma}\Psi_{\tau}\bar{\Psi}_{\tau}\bar{\Xi}_{\sigma}}$ as the initial tensor, to obtain the coarse-grained Grassmann tensor $\mathcal{T}_{\Xi'_{\sigma}\Psi'_{\tau}\bar{\Psi}'_{\tau}\bar{\Xi}'_{\sigma}}$. Note that with sufficiently small $\epsilon(<1)$, little truncation error is accumulated with the first $m_{\tau}$ times of renormalization along $\tau$-direction. This is because the contribution from the spatial hopping terms, which are of $O(\epsilon)$, is smaller than that from the temporal one, which is of $O(1)$. For the $(1+1)d$ Hubbard model, we found that the optimal $m_{\tau}$ satisfied the condition $\epsilon 2^{m_{\tau}}\sim O(10^{-1})$. \footnote{A similar remark is also mentioned in Ref.~\cite{PhysRevB.86.045139}, where the $3d$ HOTRG is applied to $2d$ quantum transverse Ising model in the path-integral formalism.}

When one applies the TRG approach to evaluate the path integral over the Grassmann fields, it is practically useful to encode the Grassmann parity of the auxiliary Grassmann fields into the subscripts of the coefficient tensor. We identify the coefficient tensor in Eq.~\eqref{eq:gtensor} as a four-rank tensor $T_{xtx't'}$, where $x,x'=1,\cdots,2^4$ and $t,t'=1,\cdots,2^2$. These new indices are defined as in Tables~\ref{tab:index_x} and \ref{tab:index_t}. Notice that $x(x')=1,\cdots,8$ correspond to the Grassmann-even sector and $x(x')=9,\cdots,16$ the Grassmann-odd one in $\Psi_{\sigma}(\bar{\Psi}_{\sigma})$. Similarly, $t(t')=1,2$ correspond to the Grassmann-even sector and $t(t')=3,4$ the Grassmann-odd one in $\Psi_{\tau}(\bar{\Psi}_{\tau})$. These mappings help us to carry out the singular value decompositions with some block-diagonal representations as explained in Ref.~\cite{Akiyama:2020soe}.

\begin{table}[htbp]
	\label{tab:index_x}
	\caption{Mapping of spatial subscripts.}
	\begin{center}
		\begin{tabular}{|c|cccccccc|cccccccc|}\hline
    		$x$ & 1 & 2 & 3 & 4 & 5 & 6 & 7 & 8 & 9 & 10 & 11 & 12 & 13 & 14 & 15 & 16 \\ \hline
    		$i_{\sigma,\uparrow}$      & 0 & 1 & 1 & 1 & 0 & 0 & 0 & 1 & 1 & 0 & 0 & 1 & 0 & 1 & 1 & 0\\
    		$i_{\sigma,\downarrow}$ & 0 & 1 & 0 & 0 & 1 & 1 & 0 & 1 & 0 & 1 & 0 & 1 & 0 & 1 & 0 & 1\\
    		$j_{\sigma,\uparrow}$      & 0 & 0 & 1 & 0 & 1 & 0 & 1 & 1 & 0 & 0 & 1 & 1 & 0 & 0 & 1 & 1\\
    		$j_{\sigma,\downarrow}$ & 0 & 0 & 0 & 1 & 0 & 1 & 1 & 1 & 0 & 0 & 0 & 0 & 1 & 1 & 1 & 1\\ \hline
	\end{tabular}
	\end{center}
\end{table}

\begin{table}[htbp]
	\label{tab:index_t}
	\caption{Mapping of temporal subscripts.}
	\begin{center}
		\begin{tabular}{|c|cc|cc|}\hline
    		$t$ & 1 & 2 & 3 & 4 \\ \hline
    		$i_{\tau,\uparrow}$      & 0 & 1 & 1 & 0  \\
    		$i_{\tau,\downarrow}$ & 0 & 1  & 0 & 1\\ \hline
	\end{tabular}
	\end{center}
\end{table}

\section{Numerical results} 
\label{sec:results}

\begin{figure}[htbp]
	\centering
	\includegraphics[width=0.7\hsize]{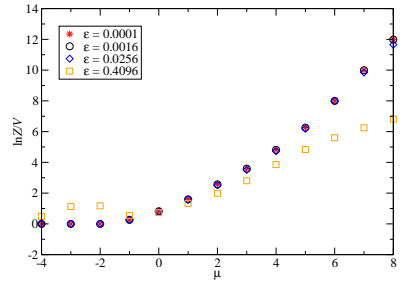}
 	\caption{Thermodynamic potential at $U/t=4$ on $V=4096\times 1677.7216$ lattice as a function of chemical potential $\mu$. $\beta$ is divided with $\epsilon=2^{12}\times10^{-4},2^{8}\times10^{-4},2^{4}\times10^{-4}$, and $10^{-4}$. The bond dimension is chosen to be $D=80$.}
  	\label{fig:lnZ_delbeta}
\end{figure}

The partition function of Eq.~\eqref{eq:Z} is evaluated using the numerical algorithm explained above on lattices with the physical volume $V=L\times \beta=N_{\sigma}\times (\epsilon N_\tau)$ ($N_{\sigma}, N_\tau=2^m, m \in \mathbb{N}$) with the periodic boundary condition for the spacial direction and the anti-periodic one for the temporal direction.
We employ  $t=1$ for the hopping parameter and $U=4$ for the four-fermi coupling. In Fig.~\ref{fig:lnZ_delbeta} we plot the $\mu$ dependence of the thermodynamic potential $\ln Z/V$ on $V=L\times\beta=4096\times 1677.7216$ with the bond dimension $D=80$ in the HOTRG algorithm choosing $\epsilon=2^{12}\times10^{-4},2^{8}\times10^{-4},2^{4}\times10^{-4},10^{-4}$. For each value of $\epsilon$, $m_{\tau}$ is decided via the condition $\epsilon 2^{m_{\tau}}=2^{12}\times10^{-4}=O(10^{-1})$. We find clear discretization effects for the $\epsilon=2^{12}\times10^{-4}$ case. On the other hand, the results with $\epsilon=2^{4}\times10^{-4}$ and $10^{-4}$ show good consistency. This means that the discretization effects with $\epsilon=10^{-4}$ are negligible. 

\begin{figure}[htbp]
	\centering
	\includegraphics[width=0.7\hsize]{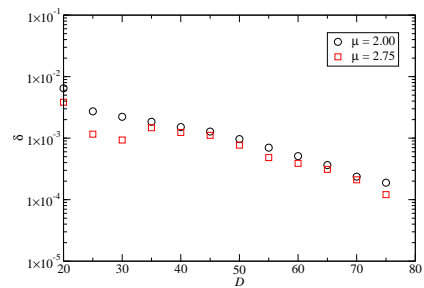}
	\caption{Convergence behavior of thermodynamic potential with $\delta$ of Eq.~\eqref{eq:delta} at $\mu=2.00$ and 2.75 as a function of $D$ on $V=4096\times 1677.7216$ lattice.}
  	\label{fig:lnZ_D}
\end{figure}

\begin{figure}[htbp]
	\centering
	\includegraphics[width=0.7\hsize]{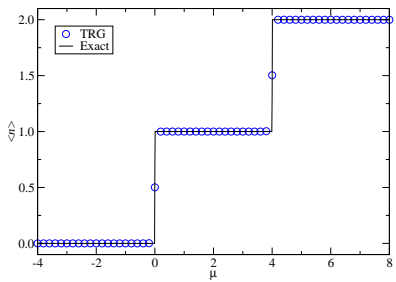}
	\caption{Electron density $\langle n\rangle$ in the $(U,t)=(4,0)$ case at $\beta=1677.7216$ with $\epsilon=10^{-4}$ as a function of $\mu$. The solid line shows the exact solution and the blue circles are the results obtained by the TRG approach.}
  	\label{fig:edensity_t0}
\end{figure}

We investigate the convergence behavior of the thermodynamic potential defining the quantity
\begin{align}
	\delta=\left|\frac{\ln Z(D)-\ln Z(D=80)}{\ln Z(D=80)}\right|
\label{eq:delta}
\end{align}
on $V=4096\times 1677.7216$ lattice with $\epsilon=10^{-4}$. In Fig.~\ref{fig:lnZ_D}, we plot the $D$ dependence of $\delta$  at $\mu=2.75$ and 2.00, which are near and far away from the critical point $\mu_{\rm c}$, respectively, as we will see below. We observe that $\delta$ decreases as a function of $D$ and reaches  $O(10^{-4})$ at $D=75$ for both values of $\mu$. Hereafter we present the results with $D=80$ except Fig.~\ref{fig:edensity_t0}.

\begin{figure}[htbp]
  	\centering
	\includegraphics[width=0.7\hsize]{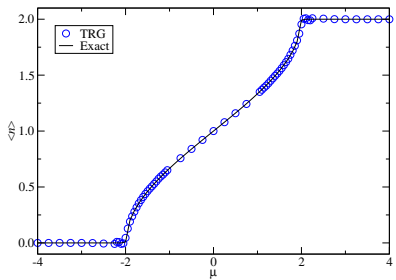}
	\caption{Electron density $\langle n\rangle$ in the $(U,t)=(0,1)$ case at $N_{\sigma}=4096$ and $\beta=1677.7216$ with $\epsilon=10^{-4}$ as a function of $\mu$. The solid line shows the exact solution on $N_{\sigma}=4096$ and the blue circles are the results obtained by the TRG approach with $D=80$.}
  	\label{fig:edensity_U0}
\end{figure}

Before presenting the $U/t=4$ results let us consider the $(U,t)=(4,0)$ and $(0,1)$ cases. Since these cases are analytically solvable, it is instructive to compare the numerical results for the electron density with the exact ones.
The electron density $\langle n\rangle$ is obtained by the numerical derivative of the thermodynamic potential in terms of $\mu$:
\begin{align}
	\langle n\rangle=\frac{1}{V}\frac{\partial \ln Z(\mu)}{\partial \mu}\approx
	\frac{1}{V}\frac{\ln Z(\mu+\Delta \mu)-\ln Z(\mu-\Delta \mu)}{2\Delta \mu}.
\end{align}
In Figs.~\ref{fig:edensity_t0} and \ref{fig:edensity_U0} we compare the numerical and analytic results for the $\mu$ dependence of $\langle n\rangle$. In both cases we observe good consistencies over the wide range of $\mu$. Note that for the case of $(U,t)=(4,0)$ in Fig.~\ref{fig:edensity_t0}, we set $m_{\tau}=24$ because this case is equivalent to the model defined on $V=1\times\beta$ lattice. Thanks to the vanishing hopping structure in the spatial direction, we can always perform an exact tensor contraction in Eq.~\eqref{eq:gtn}. In Fig.~\ref{fig:edensity_U0} we employ finer resolution of $\mu$ around $1\lesssim|\mu|\lesssim2$ in order to follow the complicated $\mu$ dependence of $\langle n\rangle$.

\begin{figure}[htbp]
  	\centering
	\includegraphics[width=0.7\hsize]{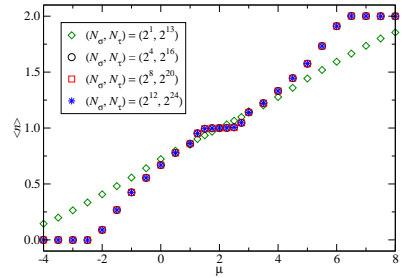}
	\caption{Electron density $\langle n\rangle$ at several lattice sizes with $\epsilon=10^{-4}$ as a function of $\mu$. The bond dimension is chosen to be $D=80$.}
  	\label{fig:edensity_beta}
\end{figure}

\begin{figure}[htbp]
  	\centering
	\includegraphics[width=0.7\hsize]{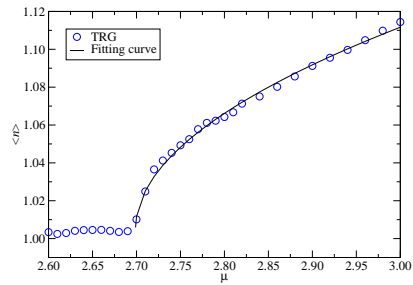}
	\caption{Electron density $\langle n\rangle$ at $\beta=1677.7216$ with $\epsilon=10^{-4}$ as a function of $\mu$. The bond dimension is chosen to be $D=80$.}
  	\label{fig:edensity}
\end{figure}

Now let us turn to the $(U,t)=(4,1)$ case.
Figure~\ref{fig:edensity_beta} shows the lattice size dependence of $\langle n\rangle$ with $\epsilon=10^{-4}$ and $m_{\tau}=12$. The results indicate that the size $(N_{\sigma},N_{\tau})=(2^{12},2^{24})$, which corresponds to $V=4096\times 1677.7216$, is sufficiently large to be identified as the thermodynamic and zero-temperature limit. The half-filling state is characterized by the plateau with $\langle n\rangle=1$ in the range of $1.3\lesssim \mu\lesssim 2.7$. We also observe the continuous change from $\langle n\rangle=1$ to $\langle n\rangle=2$ over the range of $2.7\lesssim \mu\lesssim 6.5$.
Figure~\ref{fig:edensity} shows $\mu$ dependence of $\langle n\rangle$ near the criticality on $V=4096\times 1677.7216$. The abrupt change of $\langle n\rangle$ at $\mu\approx 2.70$ in Fig.~\ref{fig:edensity} indicates a metal-insulator transition. 

We determine the critical chemical potential $\mu_{\rm c}(D)$ and the critical exponent $\nu$ on $V=4096\times 1677.7216$ lattice by fitting $\langle n\rangle$ in the metallic phase around the transition point with the following form:
\begin{align}
	\langle n\rangle=A+B\left|\mu-\mu_{\rm c}(D)\right|^\nu,
\end{align}
where $A$, $B$, $\mu_{\rm c}(D)$ and $\nu$ are the fit parameters.
The solid curve in Fig.~\ref{fig:edensity} shows the fitting result over the range of $2.68\le \mu\le 3.00$. We obtain $\mu_{\rm c}(D)=2.698(1)$ and $\nu= 0.51(2)$ at $D=80$.  Our result for the critical exponent is consistent with the theoretical prediction of $\nu=1/2$. A previous Quantum Monte Carlo simulation with small spatial extension up to $L=24$ also yielded the same conclusion~\cite{PhysRevLett.76.3176}.

\begin{table*}[htb]
	\label{tab:mu_c}
	\caption{Critical chemical potential $\mu_{\rm c}(D)$ and critical exponent $\nu$ at each $D$.}
	\begin{center}
		\begin{tabular}{|c|cccccc|}\hline
    		$D$ & 60 & 65 & 70 & 75 & 80 & $\infty$ \\ 
    		{\rm fit\; range} & [2.72,3.00] & [2.70,3.00] & [2.70,3.00] & [2.69,3.00] & [2.68,3.00] & $-$ \\ \hline
    		$\mu_{\rm c}(D)$ & 2.720(3) & 2.710(1) & 2.7068(8) & 2.701(1) & 2.698(1) & 2.642(05)(13)\\ 
    		$\nu$ & 0.49(3) & 0.52(1) & 0.50(2) & 0.51(2) & 0.51(2) & $-$\\ \hline
	\end{tabular}
	\end{center}
\end{table*}

\begin{figure}[htbp]
	\centering
	\includegraphics[width=0.7\hsize]{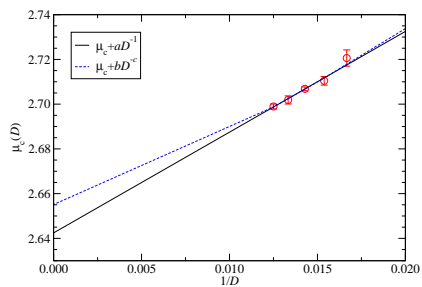}
	\caption{Critical chemical potential $\mu_{\rm c}(D)$ as a function of $1/D$. Solid line represents the fitting result with the function $\mu_{\rm c}(D)=\mu_{\rm c}+aD^{-1}$. Dotted curve also shows the fitting result with the function $\mu_{\rm c}(D)=\mu_{\rm c}+bD^{-c}$.}
 	\label{fig:mu_c}
\end{figure}

In order to extrapolate the result of $\mu_{\rm c}(D)$ to the limit $D\to\infty$, we repeat the calculation changing $D$. The numerical results are summarized in Table~\ref{tab:mu_c}. In Fig.~\ref{fig:mu_c}, we plot $\mu_{\rm c}(D)$ as a function of $1/D$, providing two types of fittings. The solid line shows the fitting result with the function $\mu_{\rm c}(D)=\mu_{\rm c}+aD^{-1}$, which gives us $\mu_{\rm c}=2.642(5)$ and $a=4.5(4)$ with $\chi^2/{\rm d.o.f}=0.447093$. We have also fitted the data with the function $\mu_{\rm c}(D)=\mu_{\rm c}+bD^{-c}$, shown as the dotted curve in Fig.~\ref{fig:mu_c}, to estimate an uncertainty in the choice of the fitting function. The difference between the central values of $\mu_{\rm c}$ obtained by these two types of fittings is considered to be a systematic error. Finally, we obtain $\mu_{\rm c}=2.642(05)(13)$ as the value of $\lim_{D\to\infty}\mu_{\rm c}(D)$, which shows good consistency with the exact solution of $\mu_{\rm c}=2.643\cdots$ based on the Bethe ansatz~\cite{PhysRevLett.20.1445,LIEB20031}.

\section{Summary and outlook} 
\label{sec:summary}

We have investigated the metal-insulator transition of the (1+1)$d$ Hubbard model in the path-integral formalism employing the TRG method. Extrapolating $\mu_{\rm c}(D)$ to the limit $D\to\infty$, we have estimated the critical chemical potential, which shows good consistency with the theoretical prediction based on the Bethe ansatz. We have determined the critical exponent $\nu$, which is also consistent with the exact solution. These encouraging results show the effectiveness of the TRG approach for the study of the Hubbard model and the related fermion models being free from the sign problem. It is worth emphasizing that the TRG approach is efficient not only in the lower-dimensional systems but also in the higher-dimensional ones, as confirmed in the earlier works \cite{PhysRevB.86.045139,Wang_2014,Sakai:2017jwp,Yoshimura:2017jpk,Kuramashi:2018mmi,Akiyama:2019xzy,Adachi:2019paf,Kadoh:2019kqk,Akiyama:2020ntf,Akiyama:2020soe,Akiyama:2021zhf}. As a next step, we are planning to investigate the phase diagram of the higher-dimensional Hubbard models, improving the TRG method successfully applied in this work.

\begin{acknowledgments}
Numerical calculation for the present work was carried out with the Oakforest-PACS (OFP) under the Interdisciplinary Computational Science Program of Center for Computational Sciences, University of Tsukuba. This work is supported in part by Grants-in-Aid for Scientific Research from the Ministry of Education, Culture, Sports, Science and Technology (MEXT) (No. 20H00148) and JSPS KAKENHI Grant Number JP21J11226 (S.A.).
\end{acknowledgments}

\appendix

\section{Grassmann tensor for $(d+1)$-dimensional Hubbard model}
\label{app:gtensor}

In this appendix, we consider the tensor network representation for the path integral of the $(d+1)$-dimensional Hubbard model, whose action is given by
\begin{widetext}
\begin{align}
\label{eq:d+1_action}
        S&=\sum_{n\in\Lambda_{d+1}}\epsilon\left\{\bar{\psi}(n)\left(\frac{\psi(n+{\hat \tau})-\psi(n)}{\epsilon}\right)-t\sum_{\sigma=1}^{d}\left(\bar{\psi}(n+{\hat\sigma})\psi(n)+\bar{\psi}(n)\psi(n+{\hat\sigma})\right)
         +\frac{U}{2}\left(\bar{\psi}(n)\psi(n)\right)^2-\mu\bar{\psi}(n)\psi(n)\right\},
\end{align}
\end{widetext}
where $n=((n_{\sigma})_{\sigma=1,\cdots,d},n_{\tau})\in\Lambda_{d+1}$, which denotes the $(d+1)$-dimensional anisotropic lattice.
Since the hopping terms in Eq.~\eqref{eq:d+1_action} are all diagonal in the internal space, we can immediately have the following decompositions,
\begin{widetext}
\begin{align}
\label{eq:i_sigma}
	{\rm e}^{t\epsilon\bar{\psi}(n)\psi(n+\hat{\sigma})}=\prod_{s=\uparrow,\downarrow}\int{\rm d}\bar{\eta}_{\sigma,s}(n){\rm d}\eta_{\tau,s}(n)~{\rm e}^{-\bar{\eta}_{\sigma,s}(n)\eta_{\sigma,s}(n)}\exp\left[\sqrt{t\epsilon}\bar{\psi}_{s}(n)\eta_{\sigma,s}(n)+\sqrt{t\epsilon}\bar{\eta}_{\sigma,s}(n)\psi_{s}(n+\hat{\sigma})\right],
\end{align}
\begin{align}
\label{eq:j_sigma}
	{\rm e}^{t\epsilon\bar{\psi}(n+\hat{\sigma})\psi(n)}=\prod_{s=\uparrow,\downarrow}\int{\rm d}\bar{\zeta}_{\sigma,s}(n){\rm d}\zeta_{\sigma,s}(n)~{\rm e}^{-\bar{\zeta}_{\sigma,s}(n)\zeta_{\sigma,s}(n)}\exp\left[-\sqrt{t\epsilon}\bar{\psi}_{s}(n+\hat{\sigma})\bar{\zeta}_{\sigma,s}(n)+\sqrt{t\epsilon}\zeta_{\sigma,s}(n)\psi_{s}(n)\right],
\end{align}
\begin{align}
\label{eq:i_tau}
	{\rm e}^{-\bar{\psi}(n)\psi(n+\hat{\tau})}=\prod_{s=\uparrow,\downarrow}\int{\rm d}\bar{\eta}_{\tau,s}(n){\rm d}\eta_{\tau,s}(n)~{\rm e}^{-\bar{\eta}_{\tau,s}(n)\eta_{\tau,s}(n)}\exp\left[-\bar{\psi}_{s}(n)\eta_{\tau,s}(n)+\bar{\eta}_{\tau,s}(n)\psi_{s}(n+\hat{\tau})\right].
\end{align}
\end{widetext}
One can now easily integrate out $\psi$ and $\bar{\psi}$ at the each site $n\in\Lambda_{d+1}$ independently and this defines the Grassmann tensor,
\begin{widetext}
\begin{align}
\label{eq:gintegral}
	&\mathcal{T}_{\Psi_{1}(n)\cdots\Psi_{d}(n)\Psi_{\tau}(n)\bar{\Psi}_{\tau}(n-\hat{\tau})\bar{\Psi}_{d}(n-\hat{d})\cdots\bar{\Psi}_{1}(n-\hat{1})}=\int\left(\prod_{s=\uparrow,\downarrow}{\rm d}\bar{\psi}_{s}{\rm d}\psi_{s}\right)~{\rm e}^{-U\epsilon\bar{\psi}_{\uparrow}\psi_{\uparrow}\bar{\psi}_{\downarrow}\psi_{\downarrow}+(\mu\epsilon+1)\bar{\psi}_{\uparrow}\psi_{\uparrow}+(\mu\epsilon+1)\bar{\psi}_{\downarrow}\psi_{\downarrow}}\nonumber\\
	&\times~\exp\left[\sum_{\sigma=1}^{d}\sum_{s=\uparrow,\downarrow}\left\{-\sqrt{t\epsilon}\bar{\psi}_{s}\bar{\zeta}_{\sigma,s}(n-\hat{\sigma})+\sqrt{t\epsilon}\zeta_{\sigma,s}(n)\psi_{s}\right\}\right]~\exp\left[\sum_{\sigma=1}^{d}\sum_{s=\uparrow,\downarrow}\left\{\sqrt{t\epsilon}\bar{\psi}_{s}\eta_{\sigma,s}(n)+\sqrt{t\epsilon}\bar{\eta}_{\sigma,s}(n-\hat{\sigma})\psi_{s}\right\}\right]\nonumber\\
	&\times~\exp\left[\sum_{s=\uparrow,\downarrow}\left\{-\bar{\psi}_{s}\eta_{\tau,s}(n)+\bar{\eta}_{\tau,s}(n-\hat{\tau})\psi_{s}\right\}\right],
\end{align}
\end{widetext}
with $\Psi_{\sigma}=(\eta_{\sigma,\uparrow},\eta_{\sigma,\downarrow},\zeta_{\sigma,\uparrow},\zeta_{\sigma,\downarrow})$, $\bar{\Psi}_{\sigma}=(\bar{\zeta}_{\sigma,\downarrow},\bar{\zeta}_{\sigma,\uparrow},\bar{\eta}_{\sigma,\downarrow},\bar{\eta}_{\sigma,\uparrow})$, $\Psi_{\tau}=(\eta_{\tau,\uparrow},\eta_{\tau,\downarrow})$, and $\bar{\Psi}_{\tau}=(\bar{\eta}_{\tau,\downarrow},\bar{\eta}_{\tau,\uparrow})$.
Using this Grassmann tensor $\mathcal{T}$, one obtains the tensor network representation for the path integral $Z$ of the $(d+1)$-dimensional Hubbard model as
\begin{widetext}
\begin{align}
	Z&=\int\left(\prod_{n\in\Lambda_{d+1}}{\rm d}\bar{\Psi}_{\tau}(n){\rm d}\Psi_{\tau}(n)~{\rm e}^{-\bar{\Psi}_{\tau}(n)\Psi_{\tau}(n)}\prod_{\sigma=1}^{d}{\rm d}\bar{\Psi}_{\sigma}(n){\rm d}\Psi_{\sigma}(n)~{\rm e}^{-\bar{\Psi}_{\sigma}(n)\Psi_{\sigma}(n)}\right)\nonumber\\
	&\quad\times\prod_{n\in\Lambda_{d+1}}\mathcal{T}_{\Psi_{1}(n)\cdots\Psi_{d}(n)\Psi_{\tau}(n)\bar{\Psi}_{\tau}(n-\hat{\tau})\bar{\Psi}_{d}(n-\hat{d})\cdots\bar{\Psi}_{1}(n-\hat{1})}.
\end{align}
\end{widetext}
Let us now carry out the integration over $\psi$ and $\bar{\psi}$ in Eq.~\eqref{eq:gintegral}. One finds the expression,
\begin{widetext}
\begin{align}
\label{eq:d+1_pre_gtensor}
	&\mathcal{T}_{\Psi_{1}\cdots\Psi_{d}\Psi_{\tau}\bar{\Psi}_{\tau}\bar{\Psi}_{d}\cdots\bar{\Psi}_{1}}\nonumber\\
	=&~\left(\prod_{\sigma=1}^{d}\sum_{i_{\sigma,\uparrow},i_{\sigma,\downarrow},j_{\sigma,\uparrow},j_{\sigma,\downarrow}}\right)~\sum_{i_{\tau,\uparrow},i_{\tau,\downarrow}}~\left(\prod_{\sigma=1}^{d}\sum_{i'_{\sigma,\uparrow},i'_{\sigma,\downarrow},j'_{\sigma,\uparrow},j'_{\sigma,\downarrow}}\right)~\sum_{i'_{\tau,\uparrow},i'_{\tau,\downarrow}}\nonumber\\
	\times&~(-1)^{\sum_{\sigma,s}i_{\sigma,s}}(\sqrt{t\epsilon})^{\sum_{\sigma,s}(i_{\sigma,s}+j_{\sigma,s}+i'_{\sigma,s}+j'_{\sigma,s})}\nonumber\\
	\times&~\left[\delta_{1,i_{\tau,\downarrow}+\sum_{\sigma}(i_{\sigma,\downarrow}+j'_{\sigma,\downarrow})}\delta_{1,i'_{\tau,\downarrow}+\sum_{\sigma}(i'_{\sigma,\downarrow}+j_{\sigma,\downarrow})}\delta_{1,i_{\tau,\uparrow}+\sum_{\sigma}(i_{\sigma,\uparrow}+j'_{\sigma,\uparrow})}\delta_{1,i'_{\tau,\uparrow}+\sum_{\sigma}(i'_{\sigma,\uparrow}+j_{\sigma,\uparrow})}\right.\nonumber\\
	-&~(\mu\epsilon+1)\delta_{0,i_{\tau,\downarrow}+\sum_{\sigma}(i_{\sigma,\downarrow}+j'_{\sigma,\downarrow})}\delta_{0,i'_{\tau,\downarrow}+\sum_{\sigma}(i'_{\sigma,\downarrow}+j_{\sigma,\downarrow})}\delta_{1,i_{\tau,\uparrow}+\sum_{\sigma}(i_{\sigma,\uparrow}+j'_{\sigma,\uparrow})}\delta_{1,i'_{\tau,\uparrow}+\sum_{\sigma}(i'_{\sigma,\uparrow}+j_{\sigma,\uparrow})}\nonumber\\
	-&~(\mu\epsilon+1)\delta_{1,i_{\tau,\downarrow}+\sum_{\sigma}(i_{\sigma,\downarrow}+j'_{\sigma,\downarrow})}\delta_{1,i'_{\tau,\downarrow}+\sum_{\sigma}(i'_{\sigma,\downarrow}+j_{\sigma,\downarrow})}\delta_{0,i_{\tau,\uparrow}+\sum_{\sigma}(i_{\sigma,\uparrow}+j'_{\sigma,\uparrow})}\delta_{0,i'_{\tau,\uparrow}+\sum_{\sigma}(i'_{\sigma,\uparrow}+j_{\sigma,\uparrow})}\nonumber\\
	-&~\left.\left\{U\epsilon-(\mu\epsilon+1)^{2}\right\}\delta_{0,i_{\tau,\downarrow}+\sum_{\sigma}(i_{\sigma,\downarrow}+j'_{\sigma,\downarrow})}\delta_{0,i'_{\tau,\downarrow}+\sum_{\sigma}(i'_{\sigma,\downarrow}+j_{\sigma,\downarrow})}\delta_{0,i_{\tau,\uparrow}+\sum_{\sigma}(i_{\sigma,\uparrow}+j'_{\sigma,\uparrow})}\delta_{0,i'_{\tau,\uparrow}+\sum_{\sigma}(i'_{\sigma,\uparrow}+j_{\sigma,\uparrow})}\right]\nonumber\\
	\times&~\eta_{\tau,\uparrow}^{i_{\tau,\uparrow}}\left(\prod_{\sigma}\eta_{\sigma,\uparrow}^{i_{\sigma,\uparrow}}\bar{\zeta}_{\sigma,\uparrow}^{j'_{\sigma,\uparrow}}\right)\bar{\eta}_{\tau,\uparrow}^{i'_{\tau,\uparrow}}\left(\prod_{\sigma}\bar{\eta}_{\sigma,\uparrow}^{i'_{\sigma,\uparrow}}\zeta_{\sigma,\uparrow}^{j_{\sigma,\uparrow}}\right)
	\eta_{\tau,\downarrow}^{i_{\tau,\downarrow}}\left(\prod_{\sigma}\eta_{\sigma,\downarrow}^{i_{\sigma,\downarrow}}\bar{\zeta}_{\sigma,\downarrow}^{j'_{\sigma,\downarrow}}\right)\bar{\eta}_{\tau,\downarrow}^{i'_{\tau,\downarrow}}\left(\prod_{\sigma}\bar{\eta}_{\sigma,\downarrow}^{i'_{\sigma,\downarrow}}\zeta_{\sigma,\downarrow}^{j_{\sigma,\downarrow}}\right),
\end{align}
\end{widetext}
where we have assigned the indices $i_{\sigma,s}(n)$, $j_{\sigma,s}(n)$, and $i_{\tau,s}(n)$ as the labels of the Taylor expansion for Eq.~\eqref{eq:i_sigma}, Eq.~\eqref{eq:j_sigma}, and Eq.~\eqref{eq:i_tau}, respectively. They take just 0 or 1 because of the nilpotency of the Grassmann numbers. For simplicity, we have omitted the lattice site dependences both from the auxiliary Grassmann fields and the indices of the Taylor expansion, introducing the notation $i'_{\nu,s}(n)=i_{\nu,s}(n-\hat{\nu})$. Then we sort the auxiliary Grassmann fields in Eq.~\eqref{eq:d+1_pre_gtensor} as those in Eq.~\eqref{eq:d+1_gtensor} and the Grassmann tensor $\mathcal{T}$ is finally written as
\begin{widetext}
\begin{align}
\label{eq:d+1_gtensor}
	&\mathcal{T}_{\Psi_{1}\cdots\Psi_{d}\Psi_{\tau}\bar{\Psi}_{\tau}\bar{\Psi}_{d}\cdots\bar{\Psi}_{1}}\nonumber\\
	=&~\left(\prod_{\sigma=1}^{d}\sum_{i_{\sigma,\uparrow},i_{\sigma,\downarrow},j_{\sigma,\uparrow},j_{\sigma,\downarrow}}\right)~\sum_{i_{\tau,\uparrow},i_{\tau,\downarrow}}~\left(\prod_{\sigma=1}^{d}\sum_{i'_{\sigma,\uparrow},i'_{\sigma,\downarrow},j'_{\sigma,\uparrow},j'_{\sigma,\downarrow}}\right)~\sum_{i'_{\tau,\uparrow},i'_{\tau,\downarrow}}\nonumber\\
	\times&~ T_{(i_{1,\uparrow},i_{1,\downarrow},j_{1,\uparrow},j_{1,\downarrow})\cdots(i_{d,\uparrow},i_{d,\downarrow},j_{d,\uparrow},j_{d,\downarrow})(i_{\tau,\uparrow},i_{\tau,\downarrow})(i'_{1,\uparrow},i'_{1,\downarrow},j'_{1,\uparrow},j'_{1,\downarrow})\cdots(i'_{d,\uparrow},i'_{d,\downarrow},j'_{d,\uparrow},j'_{d,\downarrow})(i'_{\tau,\uparrow},i'_{\tau,\downarrow})}\nonumber\\
	\times&~\left(\eta_{1,\uparrow}^{i_{1,\uparrow}}\eta_{1,\downarrow}^{i_{1,\downarrow}}\zeta_{1,\uparrow}^{j_{1,\uparrow}}\zeta_{1,\downarrow}^{j_{1,\downarrow}}\right)
	\cdots
	\left(\eta_{d,\uparrow}^{i_{d,\uparrow}}\eta_{d,\downarrow}^{i_{d,\downarrow}}\zeta_{d,\uparrow}^{j_{d,\uparrow}}\zeta_{d,\downarrow}^{j_{d,\downarrow}}\right)
	\left(\eta_{\tau,\uparrow}^{i_{\tau,\uparrow}}\eta_{\tau,\downarrow}^{i_{\tau,\downarrow}}\right)\nonumber\\
	\times&~\left(\bar{\eta}_{\tau,\downarrow}^{i'_{\tau,\downarrow}}\bar{\eta}_{\tau,\uparrow}^{i'_{\tau,\uparrow}}\right)
	\left(\bar{\zeta}_{d,\downarrow}^{j'_{d,\downarrow}}\bar{\zeta}_{d,\uparrow}^{j'_{d,\uparrow}}\bar{\eta}_{d,\downarrow}^{i'_{d,\downarrow}}\bar{\eta}_{d,\uparrow}^{i'_{d,\uparrow}}\right)
	\cdots
	\left(\bar{\zeta}_{1,\downarrow}^{j'_{1,\downarrow}}\bar{\zeta}_{1,\uparrow}^{j'_{1,\uparrow}}\bar{\eta}_{1,\downarrow}^{i'_{1,\downarrow}}\bar{\eta}_{1,\uparrow}^{i'_{1,\uparrow}}\right).
\end{align}
\end{widetext}
In the above expression, the coefficients of the auxiliary Grassmann fields are identified as a multi-rank tensor $T$. When $d=1~(\sigma=1)$, the coefficient tensor $T$ is given by
\begin{widetext}
\begin{align}
	&T_{(i_{\sigma,\uparrow},i_{\sigma,\downarrow},j_{\sigma,\uparrow},j_{\sigma,\downarrow})(i_{\tau,\uparrow},i_{\tau,\downarrow})(i'_{\sigma,\uparrow},i'_{\sigma,\downarrow},j'_{\sigma,\uparrow},j'_{\sigma,\downarrow})(i'_{\tau,\uparrow},i'_{\tau,\downarrow})}\nonumber\\
	=&~(-1)^{\sum_{s}i_{\sigma,s}}(\sqrt{t\epsilon})^{\sum_{s}(i_{\sigma,s}+j_{\sigma,s}+i'_{\sigma,s}+j'_{\sigma,s})}\nonumber\\
	\times&~\left[\delta_{1,i_{\tau,\downarrow}+i_{\sigma,\downarrow}+j'_{\sigma,\downarrow}}\delta_{1,i'_{\tau,\downarrow}+i'_{\sigma,\downarrow}+j_{\sigma,\downarrow}}\delta_{1,i_{\tau,\uparrow}+i_{\sigma,\uparrow}+j'_{\sigma,\uparrow}}\delta_{1,i'_{\tau,\uparrow}+i'_{\sigma,\uparrow}+j_{\sigma,\uparrow}}\right.\nonumber\\
	-&~(\mu\epsilon+1)\delta_{0,i_{\tau,\downarrow}+i_{\sigma,\downarrow}+j'_{\sigma,\downarrow}}\delta_{0,i'_{\tau,\downarrow}+i'_{\sigma,\downarrow}+j_{\sigma,\downarrow}}\delta_{1,i_{\tau,\uparrow}+i_{\sigma,\uparrow}+j'_{\sigma,\uparrow}}\delta_{1,i'_{\tau,\uparrow}+i'_{\sigma,\uparrow}+j_{\sigma,\uparrow}}\nonumber\\
	-&~(\mu\epsilon+1)\delta_{1,i_{\tau,\downarrow}+i_{\sigma,\downarrow}+j'_{\sigma,\downarrow}}\delta_{1,i'_{\tau,\downarrow}+i'_{\sigma,\downarrow}+j_{\sigma,\downarrow}}\delta_{0,i_{\tau,\uparrow}+i_{\sigma,\uparrow}+j'_{\sigma,\uparrow}}\delta_{0,i'_{\tau,\uparrow}+i'_{\sigma,\uparrow}+j_{\sigma,\uparrow}}\nonumber\\
	-&~\left\{U\epsilon-(\mu\epsilon+1)^{2}\right\}\left.\delta_{0,i_{\tau,\downarrow}+i_{\sigma,\downarrow}+j'_{\sigma,\downarrow}}\delta_{0,i'_{\tau,\downarrow}+i'_{\sigma,\downarrow}+j_{\sigma,\downarrow}}\delta_{0,i_{\tau,\uparrow}+i_{\sigma,\uparrow}+j'_{\sigma,\uparrow}}\delta_{0,i'_{\tau,\uparrow}+i'_{\sigma,\uparrow}+j_{\sigma,\uparrow}}\right]\nonumber\\
	\times&~(-1)^{R_{(i_{\sigma,\uparrow},i_{\sigma,\downarrow},j_{\sigma,\uparrow},j_{\sigma,\downarrow})(i_{\tau,\uparrow},i_{\tau,\downarrow})(i'_{\sigma,\uparrow},i'_{\sigma,\downarrow},j'_{\sigma,\uparrow},j'_{\sigma,\downarrow})(i'_{\tau,\uparrow},i'_{\tau,\downarrow})}},
\end{align}
\end{widetext}
with
\begin{widetext}
\begin{align}
	&R_{(i_{\sigma,\uparrow},i_{\sigma,\downarrow},j_{\sigma,\uparrow},j_{\sigma,\downarrow})(i_{\tau,\uparrow},i_{\tau,\downarrow})(i'_{\sigma,\uparrow},i'_{\sigma,\downarrow},j'_{\sigma,\uparrow},j'_{\sigma,\downarrow})(i'_{\tau,\uparrow},i'_{\tau,\downarrow})}\nonumber\\
	=&~i_{\sigma,\uparrow}i_{\tau,\uparrow}+i_{\sigma,\downarrow}(i_{\tau,\uparrow}+j'_{\sigma,\uparrow}+i'_{\tau,\uparrow}+i'_{\sigma,\uparrow}+j_{\sigma,\uparrow}+i_{\tau,\downarrow})\nonumber\\
	+&~j_{\sigma,\uparrow}(i_{\tau,\uparrow}+j'_{\sigma,\uparrow}+i'_{\tau,\uparrow}+i'_{\sigma,\uparrow})+j_{\sigma,\downarrow}(i_{\tau,\uparrow}+j'_{\sigma,\uparrow}+i'_{\tau,\uparrow}+i'_{\sigma,\uparrow}+i_{\tau,\downarrow}+j'_{\sigma,\downarrow}+i'_{\tau,\downarrow}+i'_{\sigma,\downarrow})\nonumber\\
	+&~i_{\tau,\downarrow}(j'_{\sigma,\uparrow}+i'_{\tau,\uparrow}+i'_{\sigma,\uparrow})+i'_{\tau,\downarrow}(j'_{\sigma,\uparrow}+i'_{\tau,\uparrow}+i'_{\sigma,\uparrow}+j'_{\sigma,\downarrow})+i'_{\tau,\uparrow}j'_{\sigma,\uparrow}+j'_{\sigma,\downarrow}(j'_{\sigma,\uparrow}+i'_{\sigma,\uparrow})+i'_{\sigma,\downarrow}i'_{\sigma,\uparrow}.
\end{align}
\end{widetext}

\bibliography{bib/formulation,bib/algorithm,bib/discrete,bib/grassmann,bib/continuous,bib/gauge,bib/review,bib/for_this_paper}

\begin{thebibliography}{26}%
\makeatletter
\providecommand \@ifxundefined [1]{%
 \@ifx{#1\undefined}
}%
\providecommand \@ifnum [1]{%
 \ifnum #1\expandafter \@firstoftwo
 \else \expandafter \@secondoftwo
 \fi
}%
\providecommand \@ifx [1]{%
 \ifx #1\expandafter \@firstoftwo
 \else \expandafter \@secondoftwo
 \fi
}%
\providecommand \natexlab [1]{#1}%
\providecommand \enquote  [1]{``#1''}%
\providecommand \bibnamefont  [1]{#1}%
\providecommand \bibfnamefont [1]{#1}%
\providecommand \citenamefont [1]{#1}%
\providecommand \href@noop [0]{\@secondoftwo}%
\providecommand \href [0]{\begingroup \@sanitize@url \@href}%
\providecommand \@href[1]{\@@startlink{#1}\@@href}%
\providecommand \@@href[1]{\endgroup#1\@@endlink}%
\providecommand \@sanitize@url [0]{\catcode `\\12\catcode `\$12\catcode
  `\&12\catcode `\#12\catcode `\^12\catcode `\_12\catcode `\%12\relax}%
\providecommand \@@startlink[1]{}%
\providecommand \@@endlink[0]{}%
\providecommand \url  [0]{\begingroup\@sanitize@url \@url }%
\providecommand \@url [1]{\endgroup\@href {#1}{\urlprefix }}%
\providecommand \urlprefix  [0]{URL }%
\providecommand \Eprint [0]{\href }%
\providecommand \doibase [0]{http://dx.doi.org/}%
\providecommand \selectlanguage [0]{\@gobble}%
\providecommand \bibinfo  [0]{\@secondoftwo}%
\providecommand \bibfield  [0]{\@secondoftwo}%
\providecommand \translation [1]{[#1]}%
\providecommand \BibitemOpen [0]{}%
\providecommand \bibitemStop [0]{}%
\providecommand \bibitemNoStop [0]{.\EOS\space}%
\providecommand \EOS [0]{\spacefactor3000\relax}%
\providecommand \BibitemShut  [1]{\csname bibitem#1\endcsname}%
\let\auto@bib@innerbib\@empty
\bibitem [{\citenamefont {Levin}\ and\ \citenamefont
  {Nave}(2007)}]{Levin:2006jai}%
  \BibitemOpen
  \bibfield  {author} {\bibinfo {author} {\bibfnamefont {M.}~\bibnamefont
  {Levin}}\ and\ \bibinfo {author} {\bibfnamefont {C.~P.}\ \bibnamefont
  {Nave}},\ }\href {\doibase 10.1103/PhysRevLett.99.120601} {\bibfield
  {journal} {\bibinfo  {journal} {Phys. Rev. Lett.}\ }\textbf {\bibinfo
  {volume} {99}},\ \bibinfo {pages} {120601} (\bibinfo {year} {2007})},\
  \Eprint {http://arxiv.org/abs/cond-mat/0611687} {arXiv:cond-mat/0611687
  [cond-mat.stat-mech]} \BibitemShut {NoStop}%
\bibitem [{\citenamefont {Xie}\ \emph {et~al.}(2012)\citenamefont {Xie},
  \citenamefont {Chen}, \citenamefont {Qin}, \citenamefont {Zhu}, \citenamefont
  {Yang},\ and\ \citenamefont {Xiang}}]{PhysRevB.86.045139}%
  \BibitemOpen
  \bibfield  {author} {\bibinfo {author} {\bibfnamefont {Z.~Y.}\ \bibnamefont
  {Xie}}, \bibinfo {author} {\bibfnamefont {J.}~\bibnamefont {Chen}}, \bibinfo
  {author} {\bibfnamefont {M.~P.}\ \bibnamefont {Qin}}, \bibinfo {author}
  {\bibfnamefont {J.~W.}\ \bibnamefont {Zhu}}, \bibinfo {author} {\bibfnamefont
  {L.~P.}\ \bibnamefont {Yang}}, \ and\ \bibinfo {author} {\bibfnamefont
  {T.}~\bibnamefont {Xiang}},\ }\href {\doibase 10.1103/PhysRevB.86.045139}
  {\bibfield  {journal} {\bibinfo  {journal} {Phys. Rev. B}\ }\textbf {\bibinfo
  {volume} {86}},\ \bibinfo {pages} {045139} (\bibinfo {year}
  {2012})}\BibitemShut {NoStop}%
\bibitem [{\citenamefont {Shimizu}\ and\ \citenamefont
  {Kuramashi}(2014{\natexlab{a}})}]{Shimizu:2014uva}%
  \BibitemOpen
  \bibfield  {author} {\bibinfo {author} {\bibfnamefont {Y.}~\bibnamefont
  {Shimizu}}\ and\ \bibinfo {author} {\bibfnamefont {Y.}~\bibnamefont
  {Kuramashi}},\ }\href {\doibase 10.1103/PhysRevD.90.014508} {\bibfield
  {journal} {\bibinfo  {journal} {Phys. Rev.}\ }\textbf {\bibinfo {volume}
  {D90}},\ \bibinfo {pages} {014508} (\bibinfo {year} {2014}{\natexlab{a}})},\
  \Eprint {http://arxiv.org/abs/1403.0642} {arXiv:1403.0642 [hep-lat]}
  \BibitemShut {NoStop}%
\bibitem [{\citenamefont {Sakai}\ \emph {et~al.}(2017)\citenamefont {Sakai},
  \citenamefont {Takeda},\ and\ \citenamefont {Yoshimura}}]{Sakai:2017jwp}%
  \BibitemOpen
  \bibfield  {author} {\bibinfo {author} {\bibfnamefont {R.}~\bibnamefont
  {Sakai}}, \bibinfo {author} {\bibfnamefont {S.}~\bibnamefont {Takeda}}, \
  and\ \bibinfo {author} {\bibfnamefont {Y.}~\bibnamefont {Yoshimura}},\ }\href
  {\doibase 10.1093/ptep/ptx080} {\bibfield  {journal} {\bibinfo  {journal}
  {PTEP}\ }\textbf {\bibinfo {volume} {2017}},\ \bibinfo {pages} {063B07}
  (\bibinfo {year} {2017})},\ \Eprint {http://arxiv.org/abs/1705.07764}
  {arXiv:1705.07764 [hep-lat]} \BibitemShut {NoStop}%
\bibitem [{\citenamefont {Adachi}\ \emph {et~al.}(2020)\citenamefont {Adachi},
  \citenamefont {Okubo},\ and\ \citenamefont {Todo}}]{Adachi:2019paf}%
  \BibitemOpen
  \bibfield  {author} {\bibinfo {author} {\bibfnamefont {D.}~\bibnamefont
  {Adachi}}, \bibinfo {author} {\bibfnamefont {T.}~\bibnamefont {Okubo}}, \
  and\ \bibinfo {author} {\bibfnamefont {S.}~\bibnamefont {Todo}},\ }\href
  {\doibase 10.1103/PhysRevB.102.054432} {\bibfield  {journal} {\bibinfo
  {journal} {Phys. Rev. B}\ }\textbf {\bibinfo {volume} {102}},\ \bibinfo
  {pages} {054432} (\bibinfo {year} {2020})},\ \Eprint
  {http://arxiv.org/abs/1906.02007} {arXiv:1906.02007 [cond-mat.stat-mech]}
  \BibitemShut {NoStop}%
\bibitem [{\citenamefont {Kadoh}\ and\ \citenamefont
  {Nakayama}(2019)}]{Kadoh:2019kqk}%
  \BibitemOpen
  \bibfield  {author} {\bibinfo {author} {\bibfnamefont {D.}~\bibnamefont
  {Kadoh}}\ and\ \bibinfo {author} {\bibfnamefont {K.}~\bibnamefont
  {Nakayama}},\ }\href@noop {} {\  (\bibinfo {year} {2019})},\ \Eprint
  {http://arxiv.org/abs/1912.02414} {arXiv:1912.02414 [hep-lat]} \BibitemShut
  {NoStop}%
\bibitem [{\citenamefont {Akiyama}\ \emph
  {et~al.}(2021{\natexlab{a}})\citenamefont {Akiyama}, \citenamefont
  {Kuramashi}, \citenamefont {Yamashita},\ and\ \citenamefont
  {Yoshimura}}]{Akiyama:2020soe}%
  \BibitemOpen
  \bibfield  {author} {\bibinfo {author} {\bibfnamefont {S.}~\bibnamefont
  {Akiyama}}, \bibinfo {author} {\bibfnamefont {Y.}~\bibnamefont {Kuramashi}},
  \bibinfo {author} {\bibfnamefont {T.}~\bibnamefont {Yamashita}}, \ and\
  \bibinfo {author} {\bibfnamefont {Y.}~\bibnamefont {Yoshimura}},\ }\href
  {\doibase 10.1007/JHEP01(2021)121} {\bibfield  {journal} {\bibinfo  {journal}
  {JHEP}\ }\textbf {\bibinfo {volume} {01}},\ \bibinfo {pages} {121} (\bibinfo
  {year} {2021}{\natexlab{a}})},\ \Eprint {http://arxiv.org/abs/2009.11583}
  {arXiv:2009.11583 [hep-lat]} \BibitemShut {NoStop}%
\bibitem [{\citenamefont {Shimizu}\ and\ \citenamefont
  {Kuramashi}(2014{\natexlab{b}})}]{Shimizu:2014fsa}%
  \BibitemOpen
  \bibfield  {author} {\bibinfo {author} {\bibfnamefont {Y.}~\bibnamefont
  {Shimizu}}\ and\ \bibinfo {author} {\bibfnamefont {Y.}~\bibnamefont
  {Kuramashi}},\ }\href {\doibase 10.1103/PhysRevD.90.074503} {\bibfield
  {journal} {\bibinfo  {journal} {Phys. Rev.}\ }\textbf {\bibinfo {volume}
  {D90}},\ \bibinfo {pages} {074503} (\bibinfo {year} {2014}{\natexlab{b}})},\
  \Eprint {http://arxiv.org/abs/1408.0897} {arXiv:1408.0897 [hep-lat]}
  \BibitemShut {NoStop}%
\bibitem [{\citenamefont {Shimizu}\ and\ \citenamefont
  {Kuramashi}(2018)}]{Shimizu:2017onf}%
  \BibitemOpen
  \bibfield  {author} {\bibinfo {author} {\bibfnamefont {Y.}~\bibnamefont
  {Shimizu}}\ and\ \bibinfo {author} {\bibfnamefont {Y.}~\bibnamefont
  {Kuramashi}},\ }\href {\doibase 10.1103/PhysRevD.97.034502} {\bibfield
  {journal} {\bibinfo  {journal} {Phys. Rev.}\ }\textbf {\bibinfo {volume}
  {D97}},\ \bibinfo {pages} {034502} (\bibinfo {year} {2018})},\ \Eprint
  {http://arxiv.org/abs/1712.07808} {arXiv:1712.07808 [hep-lat]} \BibitemShut
  {NoStop}%
\bibitem [{\citenamefont {Takeda}\ and\ \citenamefont
  {Yoshimura}(2015)}]{Takeda:2014vwa}%
  \BibitemOpen
  \bibfield  {author} {\bibinfo {author} {\bibfnamefont {S.}~\bibnamefont
  {Takeda}}\ and\ \bibinfo {author} {\bibfnamefont {Y.}~\bibnamefont
  {Yoshimura}},\ }\href {\doibase 10.1093/ptep/ptv022} {\bibfield  {journal}
  {\bibinfo  {journal} {PTEP}\ }\textbf {\bibinfo {volume} {2015}},\ \bibinfo
  {pages} {043B01} (\bibinfo {year} {2015})},\ \Eprint
  {http://arxiv.org/abs/1412.7855} {arXiv:1412.7855 [hep-lat]} \BibitemShut
  {NoStop}%
\bibitem [{\citenamefont {Kadoh}\ \emph {et~al.}(2018)\citenamefont {Kadoh},
  \citenamefont {Kuramashi}, \citenamefont {Nakamura}, \citenamefont {Sakai},
  \citenamefont {Takeda},\ and\ \citenamefont {Yoshimura}}]{Kadoh:2018hqq}%
  \BibitemOpen
  \bibfield  {author} {\bibinfo {author} {\bibfnamefont {D.}~\bibnamefont
  {Kadoh}}, \bibinfo {author} {\bibfnamefont {Y.}~\bibnamefont {Kuramashi}},
  \bibinfo {author} {\bibfnamefont {Y.}~\bibnamefont {Nakamura}}, \bibinfo
  {author} {\bibfnamefont {R.}~\bibnamefont {Sakai}}, \bibinfo {author}
  {\bibfnamefont {S.}~\bibnamefont {Takeda}}, \ and\ \bibinfo {author}
  {\bibfnamefont {Y.}~\bibnamefont {Yoshimura}},\ }\href {\doibase
  10.1007/JHEP03(2018)141} {\bibfield  {journal} {\bibinfo  {journal} {JHEP}\
  }\textbf {\bibinfo {volume} {03}},\ \bibinfo {pages} {141} (\bibinfo {year}
  {2018})},\ \Eprint {http://arxiv.org/abs/1801.04183} {arXiv:1801.04183
  [hep-lat]} \BibitemShut {NoStop}%
\bibitem [{\citenamefont {Kadoh}\ \emph {et~al.}(2020)\citenamefont {Kadoh},
  \citenamefont {Kuramashi}, \citenamefont {Nakamura}, \citenamefont {Sakai},
  \citenamefont {Takeda},\ and\ \citenamefont {Yoshimura}}]{Kadoh:2019ube}%
  \BibitemOpen
  \bibfield  {author} {\bibinfo {author} {\bibfnamefont {D.}~\bibnamefont
  {Kadoh}}, \bibinfo {author} {\bibfnamefont {Y.}~\bibnamefont {Kuramashi}},
  \bibinfo {author} {\bibfnamefont {Y.}~\bibnamefont {Nakamura}}, \bibinfo
  {author} {\bibfnamefont {R.}~\bibnamefont {Sakai}}, \bibinfo {author}
  {\bibfnamefont {S.}~\bibnamefont {Takeda}}, \ and\ \bibinfo {author}
  {\bibfnamefont {Y.}~\bibnamefont {Yoshimura}},\ }\href {\doibase
  10.1007/JHEP02(2020)161} {\bibfield  {journal} {\bibinfo  {journal} {JHEP}\
  }\textbf {\bibinfo {volume} {02}},\ \bibinfo {pages} {161} (\bibinfo {year}
  {2020})},\ \Eprint {http://arxiv.org/abs/1912.13092} {arXiv:1912.13092
  [hep-lat]} \BibitemShut {NoStop}%
\bibitem [{\citenamefont {Kuramashi}\ and\ \citenamefont
  {Yoshimura}(2020)}]{Kuramashi:2019cgs}%
  \BibitemOpen
  \bibfield  {author} {\bibinfo {author} {\bibfnamefont {Y.}~\bibnamefont
  {Kuramashi}}\ and\ \bibinfo {author} {\bibfnamefont {Y.}~\bibnamefont
  {Yoshimura}},\ }\href {\doibase 10.1007/JHEP04(2020)089} {\bibfield
  {journal} {\bibinfo  {journal} {JHEP}\ }\textbf {\bibinfo {volume} {04}},\
  \bibinfo {pages} {089} (\bibinfo {year} {2020})},\ \Eprint
  {http://arxiv.org/abs/1911.06480} {arXiv:1911.06480 [hep-lat]} \BibitemShut
  {NoStop}%
\bibitem [{\citenamefont {Akiyama}\ \emph {et~al.}(2020)\citenamefont
  {Akiyama}, \citenamefont {Kadoh}, \citenamefont {Kuramashi}, \citenamefont
  {Yamashita},\ and\ \citenamefont {Yoshimura}}]{Akiyama:2020ntf}%
  \BibitemOpen
  \bibfield  {author} {\bibinfo {author} {\bibfnamefont {S.}~\bibnamefont
  {Akiyama}}, \bibinfo {author} {\bibfnamefont {D.}~\bibnamefont {Kadoh}},
  \bibinfo {author} {\bibfnamefont {Y.}~\bibnamefont {Kuramashi}}, \bibinfo
  {author} {\bibfnamefont {T.}~\bibnamefont {Yamashita}}, \ and\ \bibinfo
  {author} {\bibfnamefont {Y.}~\bibnamefont {Yoshimura}},\ }\href {\doibase
  10.1007/JHEP09(2020)177} {\bibfield  {journal} {\bibinfo  {journal} {JHEP}\
  }\textbf {\bibinfo {volume} {09}},\ \bibinfo {pages} {177} (\bibinfo {year}
  {2020})},\ \Eprint {http://arxiv.org/abs/2005.04645} {arXiv:2005.04645
  [hep-lat]} \BibitemShut {NoStop}%
\bibitem [{\citenamefont {Yoshimura}\ \emph {et~al.}(2018)\citenamefont
  {Yoshimura}, \citenamefont {Kuramashi}, \citenamefont {Nakamura},
  \citenamefont {Takeda},\ and\ \citenamefont {Sakai}}]{Yoshimura:2017jpk}%
  \BibitemOpen
  \bibfield  {author} {\bibinfo {author} {\bibfnamefont {Y.}~\bibnamefont
  {Yoshimura}}, \bibinfo {author} {\bibfnamefont {Y.}~\bibnamefont
  {Kuramashi}}, \bibinfo {author} {\bibfnamefont {Y.}~\bibnamefont {Nakamura}},
  \bibinfo {author} {\bibfnamefont {S.}~\bibnamefont {Takeda}}, \ and\ \bibinfo
  {author} {\bibfnamefont {R.}~\bibnamefont {Sakai}},\ }\href {\doibase
  10.1103/PhysRevD.97.054511} {\bibfield  {journal} {\bibinfo  {journal} {Phys.
  Rev.}\ }\textbf {\bibinfo {volume} {D97}},\ \bibinfo {pages} {054511}
  (\bibinfo {year} {2018})},\ \Eprint {http://arxiv.org/abs/1711.08121}
  {arXiv:1711.08121 [hep-lat]} \BibitemShut {NoStop}%
\bibitem [{\citenamefont {Lieb}\ and\ \citenamefont
  {Wu}(1968)}]{PhysRevLett.20.1445}%
  \BibitemOpen
  \bibfield  {author} {\bibinfo {author} {\bibfnamefont {E.~H.}\ \bibnamefont
  {Lieb}}\ and\ \bibinfo {author} {\bibfnamefont {F.~Y.}\ \bibnamefont {Wu}},\
  }\href {\doibase 10.1103/PhysRevLett.20.1445} {\bibfield  {journal} {\bibinfo
   {journal} {Phys. Rev. Lett.}\ }\textbf {\bibinfo {volume} {20}},\ \bibinfo
  {pages} {1445} (\bibinfo {year} {1968})}\BibitemShut {NoStop}%
\bibitem [{\citenamefont {Lieb}\ and\ \citenamefont {Wu}(2003)}]{LIEB20031}%
  \BibitemOpen
  \bibfield  {author} {\bibinfo {author} {\bibfnamefont {E.~H.}\ \bibnamefont
  {Lieb}}\ and\ \bibinfo {author} {\bibfnamefont {F.}~\bibnamefont {Wu}},\
  }\href {\doibase https://doi.org/10.1016/S0378-4371(02)01785-5} {\bibfield
  {journal} {\bibinfo  {journal} {Physica A: Statistical Mechanics and its
  Applications}\ }\textbf {\bibinfo {volume} {321}},\ \bibinfo {pages} {1}
  (\bibinfo {year} {2003})},\ \bibinfo {note} {statphys-Taiwan-2002: Lattice
  Models and Complex Systems}\BibitemShut {NoStop}%
\bibitem [{\citenamefont {Creutz}(1987)}]{Creutz:1986ky}%
  \BibitemOpen
  \bibfield  {author} {\bibinfo {author} {\bibfnamefont {M.}~\bibnamefont
  {Creutz}},\ }\href {\doibase 10.1103/PhysRevD.35.1460} {\bibfield  {journal}
  {\bibinfo  {journal} {Phys. Rev. D}\ }\textbf {\bibinfo {volume} {35}},\
  \bibinfo {pages} {1460} (\bibinfo {year} {1987})}\BibitemShut {NoStop}%
\bibitem [{\citenamefont {Trotter}(1959)}]{10.2307/2033649}%
  \BibitemOpen
  \bibfield  {author} {\bibinfo {author} {\bibfnamefont {H.~F.}\ \bibnamefont
  {Trotter}},\ }\href {http://www.jstor.org/stable/2033649} {\bibfield
  {journal} {\bibinfo  {journal} {Proceedings of the American Mathematical
  Society}\ }\textbf {\bibinfo {volume} {10}},\ \bibinfo {pages} {545}
  (\bibinfo {year} {1959})}\BibitemShut {NoStop}%
\bibitem [{\citenamefont {Suzuki}(1976)}]{Suzuki:1976be}%
  \BibitemOpen
  \bibfield  {author} {\bibinfo {author} {\bibfnamefont {M.}~\bibnamefont
  {Suzuki}},\ }\href {\doibase 10.1007/BF01609348} {\bibfield  {journal}
  {\bibinfo  {journal} {Commun. Math. Phys.}\ }\textbf {\bibinfo {volume}
  {51}},\ \bibinfo {pages} {183} (\bibinfo {year} {1976})}\BibitemShut
  {NoStop}%
\bibitem [{\citenamefont {Akiyama}\ and\ \citenamefont
  {Kadoh}(2020)}]{Akiyama:2020sfo}%
  \BibitemOpen
  \bibfield  {author} {\bibinfo {author} {\bibfnamefont {S.}~\bibnamefont
  {Akiyama}}\ and\ \bibinfo {author} {\bibfnamefont {D.}~\bibnamefont
  {Kadoh}},\ }\href@noop {} {\  (\bibinfo {year} {2020})},\ \Eprint
  {http://arxiv.org/abs/2005.07570} {arXiv:2005.07570 [hep-lat]} \BibitemShut
  {NoStop}%
\bibitem [{\citenamefont {Assaad}\ and\ \citenamefont
  {Imada}(1996)}]{PhysRevLett.76.3176}%
  \BibitemOpen
  \bibfield  {author} {\bibinfo {author} {\bibfnamefont {F.~F.}\ \bibnamefont
  {Assaad}}\ and\ \bibinfo {author} {\bibfnamefont {M.}~\bibnamefont {Imada}},\
  }\href {\doibase 10.1103/PhysRevLett.76.3176} {\bibfield  {journal} {\bibinfo
   {journal} {Phys. Rev. Lett.}\ }\textbf {\bibinfo {volume} {76}},\ \bibinfo
  {pages} {3176} (\bibinfo {year} {1996})}\BibitemShut {NoStop}%
\bibitem [{\citenamefont {Wang}\ \emph {et~al.}(2014)\citenamefont {Wang},
  \citenamefont {Xie}, \citenamefont {Chen}, \citenamefont {Normand},\ and\
  \citenamefont {Xiang}}]{Wang_2014}%
  \BibitemOpen
  \bibfield  {author} {\bibinfo {author} {\bibfnamefont {S.}~\bibnamefont
  {Wang}}, \bibinfo {author} {\bibfnamefont {Z.-Y.}\ \bibnamefont {Xie}},
  \bibinfo {author} {\bibfnamefont {J.}~\bibnamefont {Chen}}, \bibinfo {author}
  {\bibfnamefont {B.}~\bibnamefont {Normand}}, \ and\ \bibinfo {author}
  {\bibfnamefont {T.}~\bibnamefont {Xiang}},\ }\href {\doibase
  10.1088/0256-307x/31/7/070503} {\bibfield  {journal} {\bibinfo  {journal}
  {Chinese Physics Letters}\ }\textbf {\bibinfo {volume} {31}},\ \bibinfo
  {pages} {070503} (\bibinfo {year} {2014})}\BibitemShut {NoStop}%
\bibitem [{\citenamefont {Kuramashi}\ and\ \citenamefont
  {Yoshimura}(2019)}]{Kuramashi:2018mmi}%
  \BibitemOpen
  \bibfield  {author} {\bibinfo {author} {\bibfnamefont {Y.}~\bibnamefont
  {Kuramashi}}\ and\ \bibinfo {author} {\bibfnamefont {Y.}~\bibnamefont
  {Yoshimura}},\ }\href {\doibase 10.1007/JHEP08(2019)023} {\bibfield
  {journal} {\bibinfo  {journal} {JHEP}\ }\textbf {\bibinfo {volume} {08}},\
  \bibinfo {pages} {023} (\bibinfo {year} {2019})},\ \Eprint
  {http://arxiv.org/abs/1808.08025} {arXiv:1808.08025 [hep-lat]} \BibitemShut
  {NoStop}%
\bibitem [{\citenamefont {Akiyama}\ \emph {et~al.}(2019)\citenamefont
  {Akiyama}, \citenamefont {Kuramashi}, \citenamefont {Yamashita},\ and\
  \citenamefont {Yoshimura}}]{Akiyama:2019xzy}%
  \BibitemOpen
  \bibfield  {author} {\bibinfo {author} {\bibfnamefont {S.}~\bibnamefont
  {Akiyama}}, \bibinfo {author} {\bibfnamefont {Y.}~\bibnamefont {Kuramashi}},
  \bibinfo {author} {\bibfnamefont {T.}~\bibnamefont {Yamashita}}, \ and\
  \bibinfo {author} {\bibfnamefont {Y.}~\bibnamefont {Yoshimura}},\ }\href
  {\doibase 10.1103/PhysRevD.100.054510} {\bibfield  {journal} {\bibinfo
  {journal} {Phys. Rev.}\ }\textbf {\bibinfo {volume} {D100}},\ \bibinfo
  {pages} {054510} (\bibinfo {year} {2019})},\ \Eprint
  {http://arxiv.org/abs/1906.06060} {arXiv:1906.06060 [hep-lat]} \BibitemShut
  {NoStop}%
\bibitem [{\citenamefont {Akiyama}\ \emph
  {et~al.}(2021{\natexlab{b}})\citenamefont {Akiyama}, \citenamefont
  {Kuramashi},\ and\ \citenamefont {Yoshimura}}]{Akiyama:2021zhf}%
  \BibitemOpen
  \bibfield  {author} {\bibinfo {author} {\bibfnamefont {S.}~\bibnamefont
  {Akiyama}}, \bibinfo {author} {\bibfnamefont {Y.}~\bibnamefont {Kuramashi}},
  \ and\ \bibinfo {author} {\bibfnamefont {Y.}~\bibnamefont {Yoshimura}},\
  }\href@noop {} {\  (\bibinfo {year} {2021}{\natexlab{b}})},\ \Eprint
  {http://arxiv.org/abs/2101.06953} {arXiv:2101.06953 [hep-lat]} \BibitemShut
  {NoStop}%
\end{thebibliography}%

\end{document}